\documentclass[preprint,preprintnumbers,amsmath,amssymb]{revtex4}
\pdfoutput=1
\usepackage{epsfig}
\usepackage{graphicx}
\usepackage{dcolumn}
\usepackage{bm}
\usepackage{threeparttable}
\usepackage{centernot}

\def\beq{\begin{equation}}
\def\eeq{\end{equation}}
\def\eeqn{\end{equation}}
\newcommand\iden{\leavevmode\hbox{\small1\normalsize\kern-.33em1}}

\newcommand{\bea} {\begin{eqnarray}}
\newcommand{\eea} {\end{eqnarray}}

\def\lam{\lambda}

\newcommand{\nn}{\nonumber}

\let\jnfont=\rm
\def\NPB#1 {{\jnfont Nucl.\ Phys.\ B }{\bf #1} }
\def\PLB#1 {{\jnfont Phys.\ Lett.\ B }{\bf #1} }
\def\EPJC#1 {{\jnfont Eur.\ Phys.\ Jour.\ C }{\bf #1} }
\def\PRD#1 {{\jnfont Phys.\ Rev.\ D }{\bf #1} }
\def\PRL#1 {{\jnfont Phys.\ Rev.\ Lett.\ }{\bf #1} }
\def\MPLA#1 {{\jnfont Mod.\ Phys.\ Lett.\ A }{\bf #1} }
\def\JPG#1 {{\jnfont J.\ Phys.\ G }{\bf #1} }
\def\CTP#1 {{\jnfont Commun.\ Theor.\ Phys.\ }{\bf #1} }
\def\JHEP#1 {{\jnfont JHEP \ }{\bf #1} }
\def\NPPS#1 {{\jnfont Nucl.\ Phys.\ Proc.\ Suppl.\ }{\bf #1} }
\def\CPC#1 {{\jnfont Comput.\ Phys.\ Commun.\ }{\bf #1} }
\def\CPL#1 {{\jnfont Chin.\ Phys.\ Lett. }{\bf #1} }
\def\APPB#1 {{\jnfont Acta\ Phys.\ Polon.\ B }{\bf #1} }

\def\lsim{\raise0.3ex\hbox{$<$\kern-0.75em\raise-1.1ex\hbox{$\sim$}}}
\def\gsim{\raise0.3ex\hbox{$>$\kern-0.75em\raise-1.1ex\hbox{$\sim$}}}
\def\PR#1 {{\jnfont Phys.\ Rept. }{\bf #1} }
\def\CHC#1 {{\jnfont Chin.\ Phys.\ C }{\bf #1} }
\def\NIMA#1 {{\jnfont Nucl.\ Instrum.\ Meth.\ A }{\bf #1} }
\def\JCAP#1 {{\jnfont JCAP \ }{\bf #1} }
\def\ASA#1 {{\jnfont Astron.\ Astrophys.\ A }{\bf #1} }  

\begin{document}

\title{\ \\[10mm] The CDF $W$-mass, muon $g-2$, and dark matter in a $U(1)_{L_\mu-L_\tau}$ model with vector-like leptons}

\author{Quan Zhou, Xiao-Fang Han$^{*}$\footnotetext{*) 
xfhan@ytu.edu.cn}}

\affiliation{Department of Physics, Yantai University, Yantai 264005, P. R. China}


\begin{abstract}
We study the CDF $W$-mass, muon $g-2$, and dark matter observables in a local $U(1)_{L_\mu-L_\tau}$ model in
which the new particles include three vector-like leptons ($E_1,~ E_2,~ N$), a new gauge boson $Z'$, a scalar $S$ (breaking $U(1)_{L_\mu-L_\tau}$), a scalar dark matter 
$X_I$ and its partner $X_R$. We find that the CDF $W$-mass disfavors $m_{E_1}= m_{E_2}={m_N}$ or $s_L=s_R=0$ where $s_{L(R)}$ is mixing parameter of left (right)-handed fields of 
vector-like leptons. A large mass splitting between $E_1$ and $E_2$ is favored when the differences between $s_L$ and $s_R$ becomes small.
The muon $g-2$ anomaly can be simultaneously explained for appropriate difference between $s_L$ $(m_{E_1})$ and $s_R$ $(m_{E_2})$, and some regions
are excluded by the diphoton signal data of the 125 GeV Higgs. Combined with the CDF $W$-mass, muon $g-2$ anomaly and other relevant constraints, 
the correct dark matter relic density is mainly obtained in two different scenarios:
(i) $X_IX_I\to Z'Z',~ SS$ for $m_{Z'}(m_S)<m_{X_I}$ and (ii) the co-annihilation processes for $min(m_{E_1},m_{E_2},m_N,m_{X_R})$ closed to $m_{X_I}$.
Finally, we use the direct searches for $2\ell+E_T^{miss}$ event at the LHC to constrain the model, and show the allowed mass ranges of the vector-like leptons and dark matter.
\end{abstract}

\maketitle

\section{introduction}
Recently, the CDF collaboration reported their new measurement of the $W$-boson mass \cite{cdfmw}
\bea
\Delta m_W=80.4335 \pm 0.0094 {\rm GeV},
\eea 
which approximately has $7\sigma$ deviation from the Standard Model (SM) value, $80.357 \pm 0.006$  GeV\cite{smmw}.
Besides, combined with the FNAL E989 experiment measurement ~\cite{mug2-exp}, the BNL experiment of the muon anomalous magnetic moment (muon $g-2$) 
~\cite{fermig2} has an approximate $4.2\sigma$ discrepancy from the SM prediction \cite{smg2-1,smg2-2,smg2-3},
\bea
\Delta a_\mu=a_\mu^{exp}-a_\mu^{SM}=(25.1\pm5.9)\times10^{-10}.
\eea 
The two anomalies both call for new physics beyond SM. 
There have been many works explaining the CDF $W$-mass  \cite{2204.03693,2204.03796,2204.03996,2204.04183,2204.04191,2204.04202,2204.04204,2204.04286,2204.04356,2204.04514,2204.04559,2204.04770,2204.04805,
2204.04834,2204.05024,2204.05031,2204.05085,2204.05260,2204.05267,2204.05269,2204.05283,2204.05284,2204.05285,2204.05296,
2204.05302,2204.05303,2204.04688,2204.05728,2204.05760,2204.05942,2204.05962,2204.05965,2204.05975,2204.05992,2204.06327,2204.06485,
2204.06505,2204.06541,2204.07144,2204.07511,2204.10130,2204.10315,2204.08390,2204.07022,2204.07138,2204.07411,2204.08266,2204.08067,
2204.08546,2204.08568,2204.09001,2204.09024,2204.09031,2204.09585,2204.09671,2204.10338,2204.10274,2204.10156,2204.11570,2204.11755,
2204.11871,2204.11945,2204.12018,2204.12453}.

In this paper, we study the CDF $W$-mass, the muon $g-2$, and the DM observables in a local $U(1)_{L_\mu-L_\tau}$ model in which
 a singlet vector-like lepton, a doublet vector-like lepton and a complex singlet $X$ field are introduced in addition to the $U(1)_{L_\mu-L_\tau}$ gauge boson $Z'$ \cite{lu-lt} and 
a complex singlet ${\cal S}$ breaking $U(1)_{L_\mu-L_\tau}$ symmetry. 
 As the lightest component of $X$, $X_I$ is
a candidate of dark matter (DM) and its heavy partner is $X_R$. 
The gauge boson self-energy diagrams exchanging the vector-like leptons in the loop can give additional contributions
to the oblique parameters ($S,~T,~U$), and explain the CDF $W$-mass. The interactions between the vector-like leptons
and muon mediated by the $X_I~(X_R)$ can enhance the muon $g-2$ \cite{1305.3522,1312.5329,1906.11297,1605.06298,1610.06587,1711.11567,1808.06639,1807.11484,2104.03202,2101.05819,Bharadwaj:2021tgp}. 
These new particles can affect the DM relic density via the DM pair-annihilation and various co-annihilations processes.

Our work is organized as follows. In Sec. II we introduce the
model. In Sec. III and Sec. IV we 
study the W-boson mass, muon g-2 anomaly, and the DM observables imposing relevant theoretical and experimental constraints.
Finally, we give our conclusion in Sec. V.

\section{The model}
In addition to the $U(1)_{L_\mu-L_\tau}$ gauge boson $Z'$, we introduce 
a complex singlet ${\cal S}$ breaking $U(1)_{L_\mu-L_\tau}$, a complex singlet $X$, and the following
vector-like lepton fields,
\begin{equation}
E^{"}_{L,R}=\left(\begin{array}{c} N_{L,R} \\
E^"_{L,R}
\end{array}\right)\,,E^{'}_{L,R} \,.
\end{equation}
 Their quantum numbers under the gauge group 
$SU(3)_C\times SU(2)_L\times U(1)_Y\times U(1)_{L_\mu-L_\tau}$
are displaced in Table \ref{tabquantum}.

\begin{table}[t]
  \centering
  \caption{The quantum numbers of $U(1)_{L_\mu-L_\tau}$. The charge of the other field is zero.}
  \label{tabquantum}
 \begin{tabular}{ccccc}
  \hline\hline
            & SU(3)$_c$~~ & SU(2)$_L$ ~~& U(1)$_Y$ ~~& U(1)$_{X}$  \\ \hline
  $ E^"_{L,R} $    & {\bf 1}   & {\bf 2}& $-1/2$ & $1-q_x$  \\
  $ E'_{L,R} $    & {\bf 1} & {\bf 1}& $-1$ & $1-q_x$  \\
  $ X $    & {\bf 1} & {\bf 1}& $0$ & $q_x$  \\
  ${\cal S}$    & {\bf 1} & {\bf 1}& $0$ & $-2q_x$  \\ 
  $ L_\mu  $    & {\bf 1} & {\bf 2}& $-1/2$ & $1$  \\
  $ \mu_R $    & {\bf 1} & {\bf 1}& $-1$ & $1$  \\
  $ L_\tau  $    & {\bf 1} & {\bf 2}& $-1/2$ & $-1$  \\
  $ \tau_R $    & {\bf 1} & {\bf 1}& $-1$ & $-1$  \\ \hline
    \end{tabular}
\end{table}

The new Lagrangian respecting the $SU(3)_C\times SU(2)_L\times U(1)_Y\times U(1)_{L_\mu-L_\tau}$ symmetry is written as
\begin{align}
  {\cal L} &= -{1 \over 4} Z'_{\mu\nu} Z^{\prime\mu\nu}
              + g_{Z'} Z'^{\mu}(\bar{\mu}\gamma_\mu \mu + \bar{\nu}_{\mu_L}\gamma_\mu\nu_{\mu_L} - \bar{\tau}\gamma_\mu \tau - \bar{\nu}_{\tau_L}
             \gamma_\mu\nu_{\tau_L})\nonumber\\
            &+ \bar{E"} (i \centernot D ) E"+ \bar{E'} (i \centernot D ) E'+ (D_\mu X^\dagger) (D^\mu X)  + (D_\mu {\cal S}^\dagger) (D^\mu {\cal S}) \nonumber\\
            &-V  + \mathcal{L}_{\rm Y}.
\label{eq:model}
\end{align} 
Where $D_\mu$ is the covariant derivative and $g_{Z'}$ is the gauge coupling 
constant of the $U(1)_{L_\mu-L_\tau}$ group. The field strength tensor $Z'_{\mu\nu}=\partial_\mu Z'_\nu-\partial_\nu Z'_\mu$, and the kinetic mixing
 term of gauge bosons of $U(1)_{L_\mu-L_\tau}$ and $U(1)_Y$ is ignored.
$V$ and $\mathcal{L}_{\rm Y}$ indicate the scalar potential and Yukawa interactions.

The scalar potential $V$ is written as
\begin{eqnarray} \label{scalarV} V &=& -\mu_{h}^2
(H^{\dagger} H) - \mu_{S}^2 ({\cal S}^{\dagger} {\cal S}) + m_X^2 (X^{\dagger} X) + \left[\mu X^2 {\cal S} + \rm h.c.\right]\nonumber \\
&&+ \lambda_H (H^{\dagger} H)^2 +
\lambda_S ({\cal S}^{\dagger} {\cal S})^2 + \lambda_X (X^{\dagger} X)^2 + \lambda_{SX}
({\cal S}^{\dagger} {\cal S})(X^{\dagger} X) \nonumber \\
&&+ \lambda_{HS}(H^{\dagger} H)({\cal S}^{\dagger} {\cal S}) + \lambda_{HX}(H^{\dagger} H)(X^{\dagger} X),
\end{eqnarray}
where the SM Higgs doublet $H$, the complex singlet fields ${\cal S}$ and $X$ are
\begin{equation}
H=\left(\begin{array}{c} G^+ \\
\frac{1}{\sqrt{2}}\,(h_1+v_h+iG)
\end{array}\right)\,,
{\cal S}={1\over \sqrt{2}} \left( h_2+v_S+i\omega\right) \,,
X={1\over \sqrt{2}} \left( X_R+iX_I\right) \,.
\end{equation}
Here $H$ and ${\cal S}$ respectively acquire vacuum expectation values (VEVs), $v_h=246$ GeV and $v_S$, and the VEV of $X$ field is zero.
The parameters $\mu^{2}_{h}$ and $\mu^{2}_{S}$ are determined by the minimization conditions for Higgs potential,
\beq
\begin{split}
&\quad \mu_{h}^2 = \lam_H v_h^2 + {1 \over 2} \lam_{HS} v_S^2,\\
&\quad \mu_{S}^2 = \lam_S v_S^2 + {1 \over 2} \lam_{HS} v_h^2.\\
\end{split}
\label{min_cond}
\eeq
The complex scalar $X$ is split into two real scalar fields $X_R$ and $X_I$ by the $\mu$ term after the ${\cal S}$ field acquires VEV $v_S$.
Their masses are 
\begin{align}\label{eqmu}
 &m_{X_R}^2 = m_X^2 +  {1 \over 2} \lambda_{HX} v_H^2 + {1 \over 2}\lambda_{SX} v_S^2  + \sqrt{2} \mu v_S\nonumber\\
 &m_{X_I}^2 = m_X^2 +  {1 \over 2} \lambda_{HX} v_H^2 + {1 \over 2}\lambda_{SX} v_S^2  - \sqrt{2} \mu v_S.
\end{align}
Because the $X$ field has no VEV, there is a remnant
discrete $Z_2$ symmetry which makes the lightest component $X_I$ to be stable and as a candidate of DM.

The $\lambda_{HS}$ term leads to a mixing of $h_1$ and $h_2$, and their mass eigenstates $h$ and $S$ are obtained from following relation,
\begin{align} 
\left(
\begin{array}{c}
h_1 \\ h_2
\end{array}
\right)
=
\left(
\begin{array}{cc}
\cos\alpha & \sin\alpha\\
-\sin\alpha & \cos\alpha \\
\end{array}
\right)
\left(
\begin{array}{c}
h \\ S
\end{array}
\right)
\end{align} 
with $\alpha$ being the mixing angle. From the $\lambda_{HS}$ term and $\lambda_{HX}$ term, we can obtain
the 125 GeV Higgs $h$ coupling to a pair of DM. In order to escape the strong bounds of the DM indirect detection and direct detection 
experiments, we simply assume the $hX_IX_I$ coupling to be absent, namely choosing $\lambda_{HS}=0$ and $\lambda_{HX}=0$.
Thus we obtain
\beq\label{eqlamhs}
\theta=0,~~~\lambda_H=\frac{m_h^2}{2v_h^2},~~~\lambda_S=\frac{m_S^2}{2v_S^2}.
\eeq
The gauge boson $Z'$ acquires a mass after ${\cal S}$ breaks the $U(1)_{L_\mu-L_\tau}$ symmetry, 
\beq\label{eqvs} 
m_{Z'} = 2g_{Z'} \mid q_x\mid v_S.
\eeq

The Yukawa interactions with the $U(1)_{L_\mu-L_\tau}$ symmetry are given as
\begin{align}
-\mathcal{L}_{\rm Y}
=&  m_1 \overline{E'_L} E'_R  + m_2 \overline{E^"_L} E^"_R + \kappa_{1} \overline{\mu_R} X E'_L+ \kappa_{2} \overline{L_\mu} X E^"_R \nn \\
  & + \sqrt{2}y_1 \overline{E^"_L} H E'_R + \sqrt{2}y_2 \overline{E^"_R} H E'_L +\frac{\sqrt{2}m_\mu}{v} \overline{L_\mu}H \mu_R
+ {\rm h.c.},
\label{lag-yuka}
\end{align} 
where $L_\mu=\left(\nu_{\mu L},\mu_{L}\right)$.

Since the $X$ field has no VeV, there is no mixing between the vector-like leptons and the muon lepton.
However, there is a mixing between the vector-like leptons $E"$ and $E'$ after the $H$ acquires the VeV, $v_h=$ 246 GeV, and their mass 
matrix is given as   
\begin{align}
 M_E = 
\begin{pmatrix}
 m_1 & y_2 v_h \\ y_1 v_h & m_2
\end{pmatrix}.   
\end{align}
We take two unitary matrices to diagnolize the mass matrix, 
\begin{align}
 U_L = 
\begin{pmatrix}
 c_L & s_L \\ -s_L & c_L
\end{pmatrix},
\quad 
 U_R = 
\begin{pmatrix}
 c_R & s_R \\ -s_R & c_R
\end{pmatrix},
\quad
U_L^\dag M_E U_R = \mathrm{diag}\left(m_{E_1}, m_{E_2}\right),
\end{align}
where $c_{L,R}^2 + s_{L,R}^2 = 1$. The $E_1$ and $E_2$ are the 
mass eigenstates of charged vector-like leptons, and the mass of neutral vector-like lepton $N$ is
\beq
m_{N}=m_2.
\eeq

From the Eq. (\ref{lag-yuka}), we can obtain the interactions between the charged vector-like leptons and muon mediated by $X_R$ and $X_I$,
\beq
-\mathcal{L}_{\rm X} \supset \frac{1}{\sqrt{2}}(X_R+ i X_I)\left[\bar{\mu}_R (\kappa_1 c_L E_{1L} -\kappa_1 s_L E_{2L}) + \bar{\mu}_L (\kappa_2 s_R E_{1R} + \kappa_2 c_R E_{2R})\right] + h.c.~,
\eeq
and the 125 GeV Higgs interactions to the charged vector-like leptons $E_1$ and $E_2$,
\bea
-\mathcal{L}_{\rm h} &\supset& \frac{m_{E_1}(c_L^2 s_R^2 +c_R^2 s_L^2)-2m_{E_2}s_L c_L s_R c_R}{v_h}~ h\bar{E}_1E_1,\nonumber\\
&&+\frac{m_{E_2}(s_L^2 c_R^2 +c_L^2 s_R^2)-2m_{E_1}s_L c_L s_R c_R}{v_h}~ h\bar{E}_2E_2.\label{yuka-he1e1}
\eea

\section{The $S,~T,~U$ parameters, $W$-mass, and muon $g-2$}
In addition to $m_h=$ 125 GeV, $v_h=$ 246 GeV, $\lambda_{HS}=0$,
$\lambda_{HX}=0$, there are many new parameters in the model. We take
$g_{Z'}$, $q_x$, $m_{Z'}$, $\lambda_X$, $\lambda_{SX}$, $m_S$, $m_{X_R}$, $m_{X_I}$, $m_{E_1}$,
$m_{E_2}$, $s_L$, $s_R$, $\kappa_1$, and $\kappa_2$ as the input parameters, which can
be used to determine other parameters.

In order to maintain the perturbativity, we conservatively take
\bea
&&\mid\lambda_{SX}\mid\leq 4\pi,~~\mid\lambda_X\mid\leq 4\pi,\nonumber\\
&&-1\leq\kappa_1 \leq 1,~~-1\leq\kappa_2 \leq 1  .
\eea
The mixing parameters $s_L$ and $s_R$ are taken as
\beq
-\frac{1}{\sqrt{2}} \leq s_L \leq \frac{1}{\sqrt{2}}, ~~-\frac{1}{\sqrt{2}} \leq s_R \leq \frac{1}{\sqrt{2}}.
\eeq
We scan over the input mass parameters in the following ranges:
\begin{align}
 & 60 {\rm GeV} \leq m_{X_I} \leq 500 {\rm GeV},~~~ m_{X_I} \leq m_{X_R} \leq 1 {\rm TeV},\nonumber\\
 &m_{X_I} \leq m_{E_1} \leq 1 {\rm TeV},~~~m_{X_I} \leq m_{E_2} \leq 1 {\rm TeV},\nonumber\\
 & 100 {\rm GeV} \leq m_{Z'} \leq 1 {\rm TeV},~~~100 {\rm GeV} \leq m_{S} \leq 1 {\rm TeV}.
\end{align}
The mass of neutral vector-like lepton $N$ is determined by $m_{E_1}$, $m_{E_2}$, $s_L$ and $s_R$, we require $m_N>m_{X_I}$.
We choose 0 $<g_{Z'}/m_{Z'}\leq$ (550 GeV)$^{-1}$ to satisfy the bound of the neutrino trident process \cite{trident}.
 We take -2 $<q_x\leq 2$, and require $\mid g_{Z'}(1-q_x)\mid \leq 1$ and $ g_{Z'}\leq 1$ to respect the perturbativity of the $Z'$ couplings.

The tree-level stability of the potential in Eq. (\ref{scalarV}) impose the following bounds,
\begin{eqnarray}
&&\lambda_H \geq 0 \,, \quad \lambda_S \geq 0 \,,\quad \lambda_X \geq 0 \,,\quad \nonumber \\
&& \lambda_{HS} \geq - 2\sqrt{\lambda_H \,\lambda_{S}}  \,, \quad
\lambda_{HX} \geq - 2\sqrt{\lambda_H \,\lambda_{X}}  \,, \quad
\lambda_{SX} \geq - 2\sqrt{\lambda_S \,\lambda_{X}} \,, \quad\nonumber \\
&&\sqrt{\lambda_{HS}+2\sqrt{\lambda_H \,\lambda_S}}~\sqrt{\lambda_{HX}+
2\sqrt{\lambda_H \,\lambda_{X}}}
~\sqrt{\lambda_{SX}+2\sqrt{\lambda_S\,\lambda_{X}}} \nonumber \\
&&+ 2\,\sqrt{\lambda_H \lambda_S \lambda_{X}} + \lambda_{HS} \sqrt{\lambda_{X}}
+ \lambda_{HX} \sqrt{\lambda_S} + \lambda_{SX} \sqrt{\lambda_H} \geq 0 \,.
\end{eqnarray}

The $H\to \gamma\gamma$ decay can be corrected by the loops of the charged vector-like leptons $E_1$ and $E_2$.
We impose the bound of the diphoton signal strength of the 125 GeV Higgs \cite{smmw}, 
\beq
\mu_{\gamma\gamma}= 1.11^{+0.1}_{-0.09}.
\eeq

\subsection{The $S,~T,~U$ parameters and $W$-mass}
The model contains the interactions of gauge bosons and vector-like leptons, 
\bea
-\mathcal{L}_{\rm VG} =&& -e \gamma \bar{E}_{1,2} \gamma^\mu E_{1,2} + Z\bar{E}_i \gamma^\mu (L_{ij}P_L+R_{ij}P_R)E_j
+ \frac{g}{2c_W}Z\bar{N} \gamma^\mu (P_L+P_R) N \nonumber\\
&&+\frac{1}{\sqrt{2}} W^+\bar{N}\gamma^\mu [(c_L P_L+c_R P_R) E_2 + (s_L P_L+ s_R P_R) E_1] + h.c.~,
\eea
where $L_{ij}$ and $R_{ij}$ are
\bea
&&L(R)_{11}=A_1 c_{L(R)}^2 +A_2 s_{L(R)}^2,~~L(R)_{22}=A_1 s_{L(R)}^2 +A_2 c_{L(R)}^2,\nonumber\\
&&L(R)_{12}=L(R)_{21}=(A_2-A_1) s_{L(R)}c_{L(R)}
\eea
with
\beq
A_1=\frac{g}{c_W}s_W^2,~~A_2=\frac{g}{c_W}(-\frac{1}{2}+s_W^2).
\eeq
Where $s_W\equiv \sin\theta_W$ and $c_W=\sqrt{1-s_W^2}$, and $\theta_W$ is the weak mixing angle.

The gauge boson self-energy diagrams exchanging the vector-like leptons in the loop can give additional contributions
to the oblique parameters ($S,~T,~U$) \cite{stu,w-stu}, which are calculated as \cite{stu,w-stu,1305.4712}
\begin{eqnarray}
\alpha(M_Z^2) \, S & = & \frac{4 s_W^2 c_W^2}{M_Z^2} \left [\Pi^{\text{NP}}_{ZZ} (M_Z^2) - \Pi^{\text{NP}}_{ZZ} (0)
    -\Pi^{\text{NP}}_{\gamma \gamma}(M_Z^2) - \frac{c_W^2-s_W^2}{c_W
        s_W} \, \Pi^{\text{NP}}_{\gamma Z}(M_Z^2)\right] ,
      \\
\alpha(M_Z^2) \, T & = & \frac{\Pi^{\text{NP}}_{WW}(0)}{M_W^2}
        - \frac{\Pi^{\text{NP}}_{ZZ}(0)}{M_Z^2} ,\label{eq-t}
          \\
\alpha(M_Z^2) \, U & = & 4 s_W^2 \left [
            \frac{\Pi^{\text{NP}}_{WW}(M_W^2)-\Pi^{\text{NP}}_{WW}(0)}
                {M_W^2} - c_W^2 \left(
                \frac{\Pi^{\text{NP}}_{ZZ}(M_Z^2)-\Pi^{\text{NP}}_{ZZ}(0)}{M_Z^2}\right)
                    \nonumber \right .\\ & & \left . - 2 s_W c_W \,
                    \frac{\Pi^{\text{NP}}_{\gamma Z}(M_Z^2)}{M_Z^2} -
                       s_W^2 \, \frac{\Pi^{\text{NP}}_{\gamma
                            \gamma}(M_Z^2)}{M_Z^2} \right ],
\label{eq:STU}
\end{eqnarray}
where the $\Pi^{\text{NP}}$ function is given in Appendix A.

Analyzing precision electroweak data including the new CDF $W$-mass, Ref. \cite{2204.03796} gave the values of $S,~T$ and $U$,
\beq\label{fit-stu}
S=0.06\pm 0.10, ~~T=0.11\pm 0.12,~~U=0.14 \pm 0.09 
\eeq
with correlation coefficients 
\beq
\rho_{ST} = 0.9, ~~\rho_{SU} = -0.59, ~~\rho_{TU} = -0.85.
\eeq
The $W$-boson mass is given as \cite{w-stu},
\beq
\Delta m_W^2=\frac{\alpha c_W^2}{c_W^2-s_W^2}m_Z^2 (-\frac{1}{2}S+c_W^2T+\frac{c_W^2-s_W^2}{4s_W^2}U).
\eeq
We perform a global fit to the values of $S,~T,~U$, and require $\chi^2 < \chi^2_{\rm min} + 6.18$ with $\chi^2_{\rm min}$ denoting the minimum of $\chi^2$. 
These surviving samples mean to be within the $2\sigma$ range in any two-dimension plane of the
model parameters fitting to the $S,~T$, and $U$ parameters.

\begin{figure}
 \epsfig{file=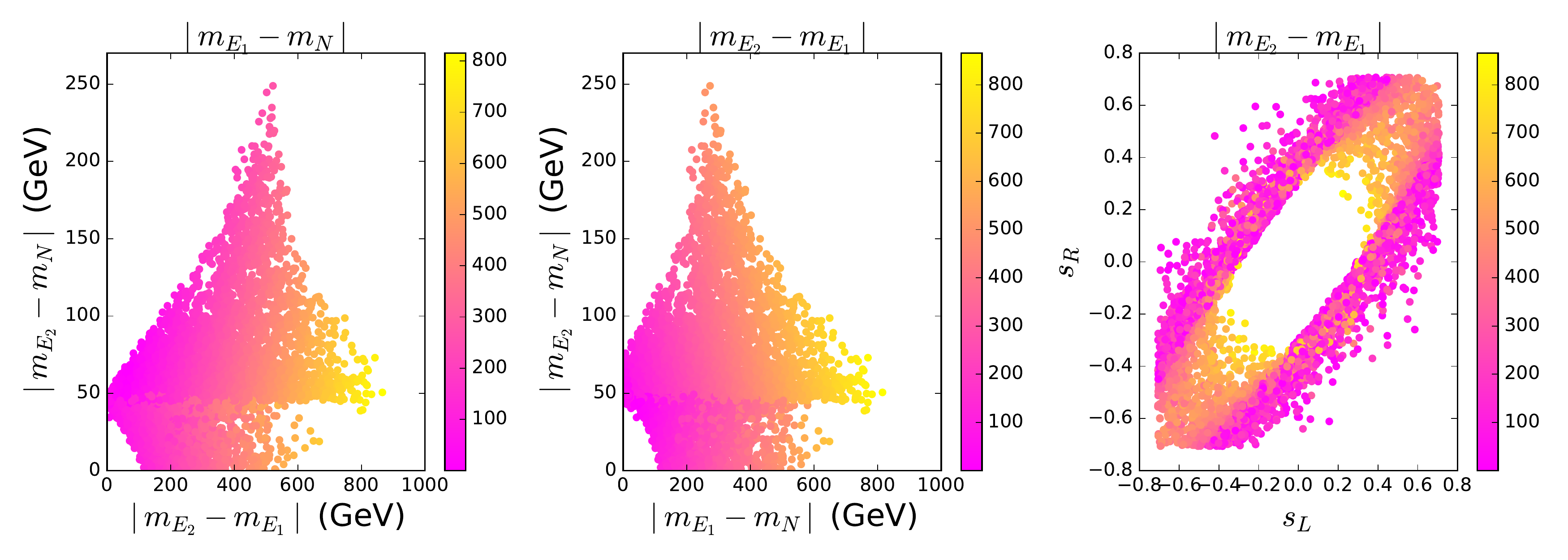,height=6.cm}
\vspace{-1.5cm} \caption{The surviving samples explaining the CDF $W$-mass within $2\sigma$ range while satisfying the oblique parameters and theoretical constraints.} \label{stumw1}
\end{figure}

In Fig. \ref{stumw1}, we show the samples explaining the CDF $W$-boson mass within $2\sigma$ range while satisfying
the constraints of the oblique parameters and theoretical constraints. Fig. \ref{stumw1} shows that the explanation of 
the CDF $W$-mass requires appropriate mass splittings among $E_1,~E_2$ and $N$, which do not simultaneously equal to zero.
For example, when $m_{E_2}=m_N$, the mass splitting between $m_{E_1}$ and $m_{E_2}(m_{N})$ is required to be larger than 100 GeV.
From the right panel of Fig. \ref{stumw1}, we see that the measurement of CDF $W$-mass disfavors $s_L$ and $s_R$ to approach to zero simultaneously,
and favors $E_1$ and $E_2$ to have a large mass splitting when the difference between $s_L$ and $s_R$ becomes small. 

Now we understand the reason. The function $\Pi^{\text{NP}}_{WW}(0)$ is zero for $m_{E_2}=m_N$ and $m_{E_1}=m_N$,
and the $\Pi^{\text{NP}}_{ZZ}(0)$ is zero for $m_{E_2}=m_{E_1}$. Therefore, from Eq. (\ref{eq-t}) we see that 
the corrections of the model to $T$ parameter are absent for $m_{E_2}=m_{E_1}=m_N$, which is disfavored by the CDF measurement of $W$ mass.
Because there is no mixing between $E_2$ and $E_1$ for $s_L=s_R=0$,  both the $ZE_2E_1$ and $WE_1N$ couplings disappear and
$m_{E_2}$ equals to $m_{N}$. Therefore, for $s_L=s_R=0$, both $\Pi^{\text{NP}}_{WW}(0)$ and $\Pi^{\text{NP}}_{ZZ}(0)$ are zero, and the corrections to $T$ parameter 
are also absent. The case of $s_L=s_R=0$ is disfavored by the CDF measurement of $W$- mass.

\subsection{The muon $g-2$}
The model can give additional corrections to the muon $g-2$ via the one-loop diagrams containing the interactions
 between muon and $E_1~ (E_2)$ mediated by $X_R$ and $X_I$, and the main corrections are calculated as \cite{1305.3522,1906.11297,0902.3360}
\bea
\Delta a_\mu &=&\frac{1}{32\pi^2}m_\mu \left(\kappa_1 c_L \kappa_2 s_R H(m_{E_1},m_{X_R})-\kappa_1 s_L \kappa_2 c_R H(m_{E_2},m_{X_R}) \nonumber \right.\\
&&\left. +\kappa_1 c_L \kappa_2 s_R H(m_{E_1},m_{X_I})-\kappa_1 s_L \kappa_2 c_R H(m_{E_2},m_{X_I}) \right),\label{eq-mug2}
\eea
where the function 
\beq
H(m_{f},m_{\phi})=\frac{m_f}{m_{\phi}^2}\frac{(r^2-4r+2\log{r}+3)}{(r-1)^3}
\eeq
with $r=\frac{m_f^2}{m_{\phi}^2}$.
Eq. (\ref{eq-mug2}) shows that the correction of the model to the muon $g-2$ is absent for $m_{E_1}=m_{E_2}$ and $s_L=s_R$. 
\begin{figure}
 \epsfig{file=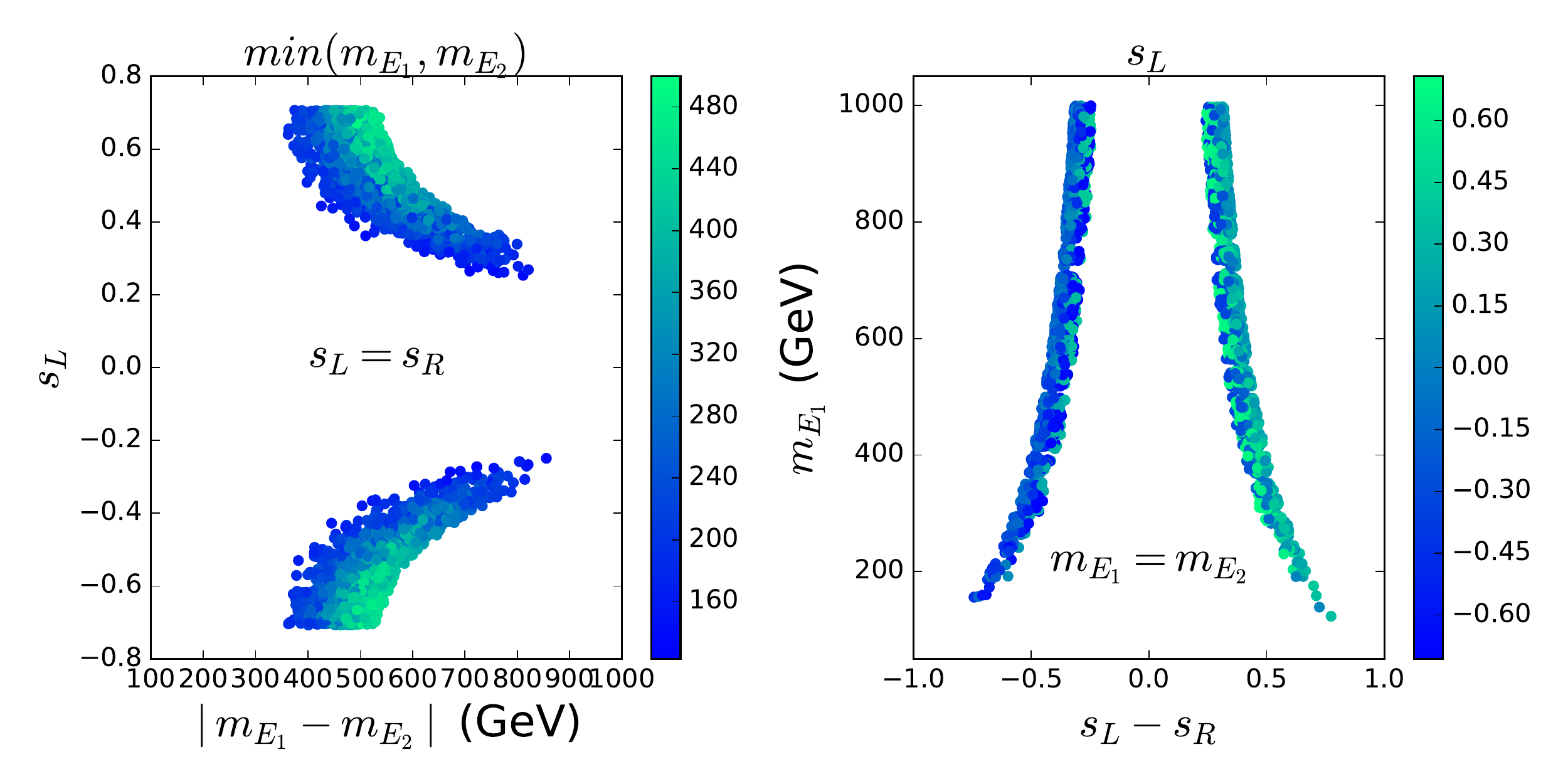,height=8.cm}
\vspace{-0.5cm} \caption{All the samples explaining the muon $g-2$ anomaly while satisfying the constraints of "pre-muon $g-2$".} \label{figg2}
\end{figure}

\begin{figure}
 \epsfig{file=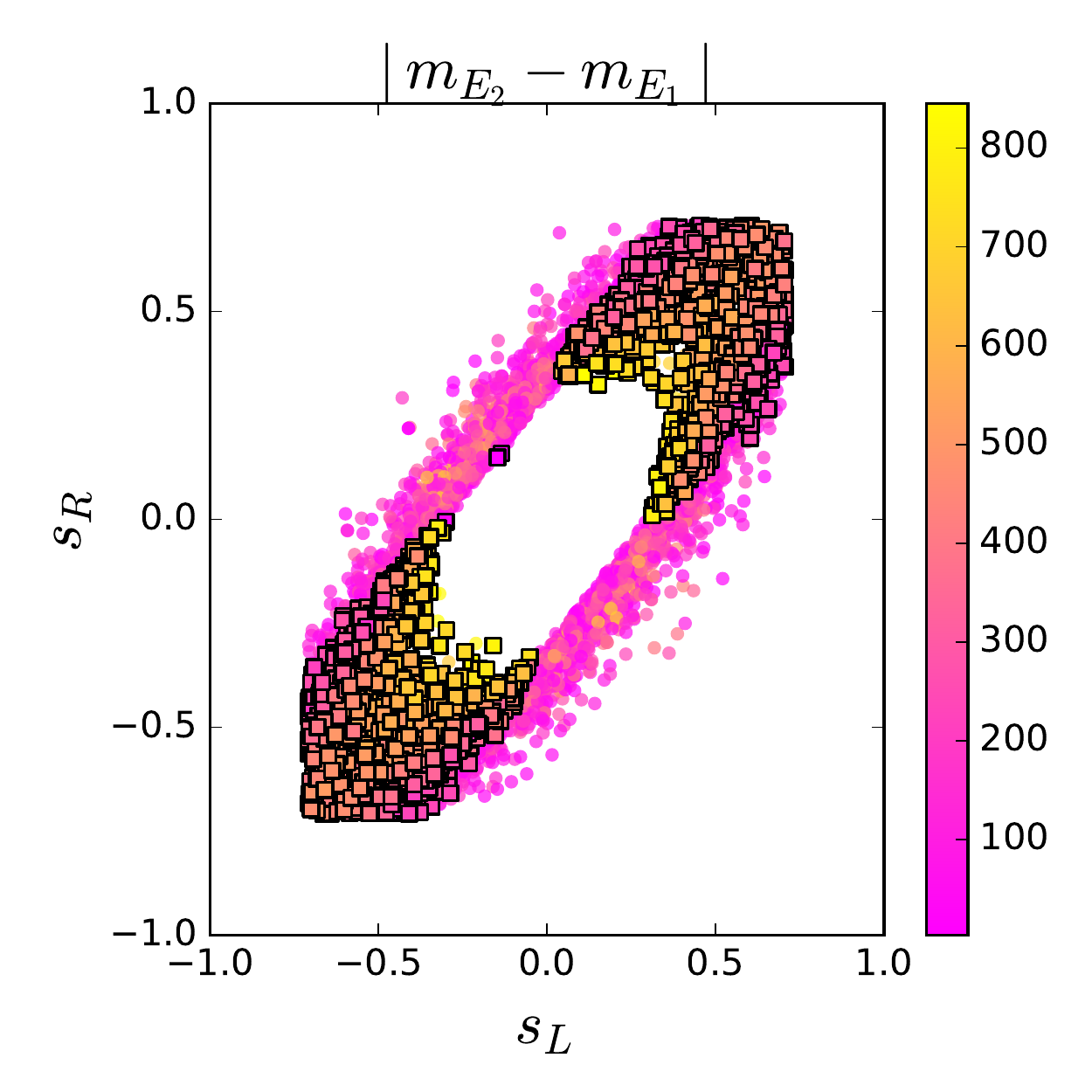,height=8.cm}
\vspace{-0.5cm} \caption{All the samples explaining the muon $g-2$ anomaly while satisfying the constraints of "pre-muon $g-2$". 
The squares and bullets are allowed and excluded by the diphoton signal data of 125 GeV Higgs, respectively.} \label{g2hrr}
\end{figure}
We respectively take $s_L=s_R$ and $m_{E_1}=m_{E_2}$, and show the samples explaining the muon $g-2$ anomaly within $2\sigma$ range while satisfying the constraints 
"pre-muon $g-2$" (denoting the theory, the oblique parameters, and the CDF $W$-mass) in Fig. \ref{figg2}. From Fig. \ref{figg2},
we see that the explanation of the muon $g-2$ anomaly favors $\mid s_L\mid$ to decrease with increasing of $\mid m_{E_1}-m_{E_2}\mid$ for $s_L=s_R$,
and $m_{E_1}$ to increase with decreasing of $\mid s_L-s_R\mid$. This characteristic can be well understood from Eq. (\ref{eq-mug2}).

\begin{figure}
 \epsfig{file=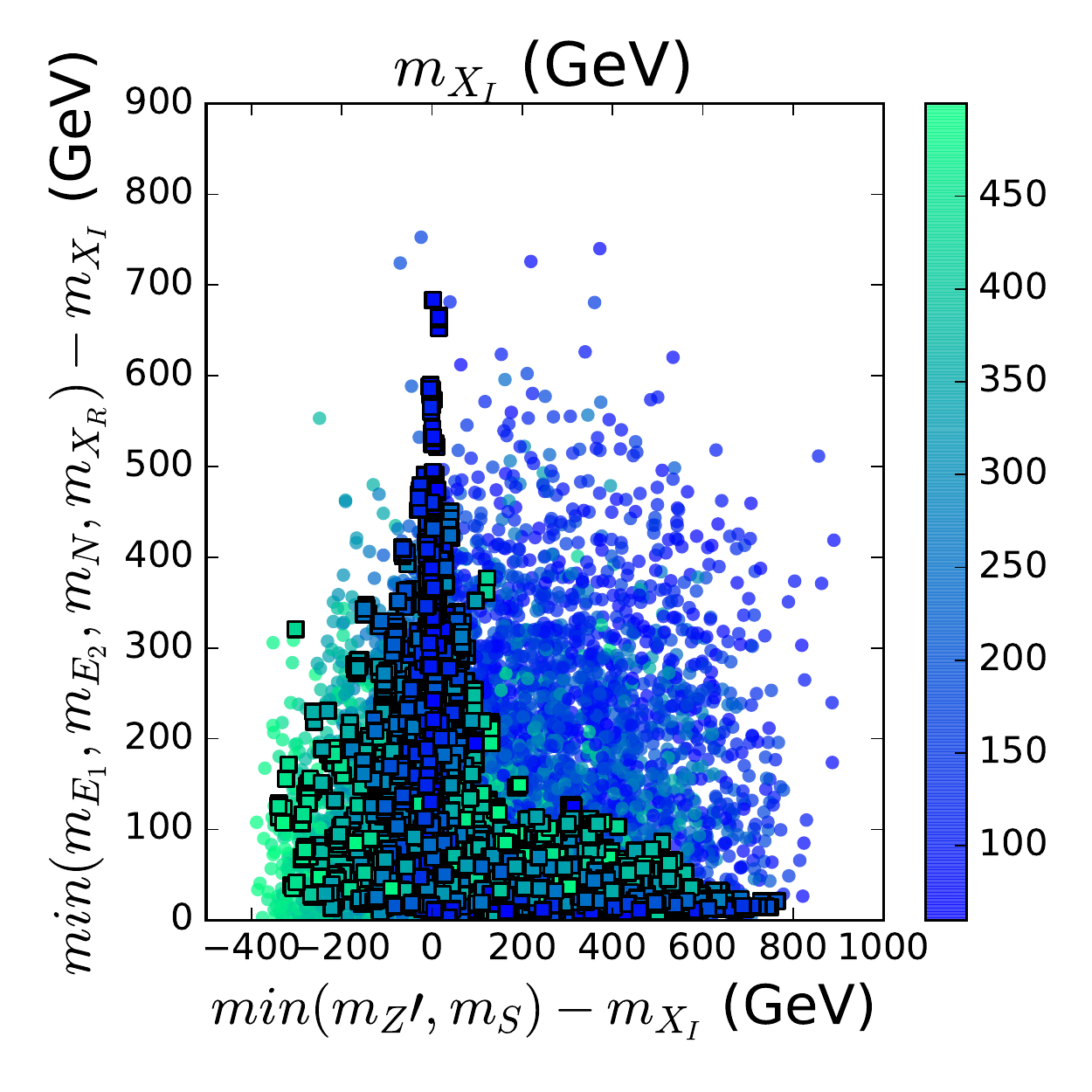,height=8.cm}
\vspace{-0.5cm} \caption{All samples satisfy the constraints of "pre-muon $g-2$", the diphoton signal data of the 125 GeV Higgs, and muon $g-2$.
The squares and bullets are allowed and excluded by the DM relic density. Here $min(m_i,m_j,\cdots)$ denotes the minimal value of $m_i,m_j,\cdots$. }
 \label{figdmall}
\end{figure}

After imposing the constraints of the diphoton signal data of the 125 GeV Higgs and "pre-muon $g-2$", we scan over the parameter space,
and project the samples explaining the muon $g-2$ anomaly in Fig. \ref{g2hrr}. We find that the diphoton signal data of the 125 GeV Higgs
exclude some samples explaining the muon $g-2$ anomaly, and favors $s_L$ and $s_R$ to have same sign, especially for large $\mid s_L\mid$ and $\mid s_R\mid$. 
When $s_L$ and $s_R$ have same sign, the terms of $h\bar{E}_1E_1$ ($h\bar{E}_2E_2$) coupling in Eq. (\ref{yuka-he1e1}) are canceled to some extent, 
which suppresses the corrections of $E_1$ and $E_2$ to the $h\to \gamma\gamma$ decay.

\section{The DM observables}
In the model, in addition to $X_IX_I\to \mu^+\mu^-$, and the DM pair-annihilation processes $X_IX_I \to Z'Z',~SS$ will be open for
$m_{Z'}~(m_S)<m_{X_I}$. When the masses of $E_1,~E_2,~N$ and $X_R$ are closed to $m_{X_I}$, their various co-annihilation processes
will play important roles in the DM relic density.
We use $\textsf{FeynRules}$ \cite{feyrule} to generate model file, and employ $\textsf{micrOMEGAs}$ \cite{micomega} to calculate the relic density.
The Planck collaboration reported the relic density of cold DM in the universe,
 $\Omega_{c}h^2 = 0.1198 \pm 0.0015$ \cite{planck}.

After imposing the constraints of "pre-muon $g-2$", the diphoton signal data of the 125 GeV Higgs, and the muon $g-2$ anomaly, we project
the samples achieving the DM relic density within $2\sigma$ range in Fig. \ref{figdmall}. Due to the constraints of muon $g-2$ on
the interactions between the vector-like leptons and muon mediated by $X_I$, it is not easy to obtain the correct DM relic density only via the $X_IX_I \to \mu^+\mu^-$ 
annihilation process, and other processes are needed to accelerate the DM annihilation. As shown in Fig. \ref{figdmall}, for $min(m_{Z'},m_S) < m_{X_I}$,
the $X_I X_I \to Z'Z'$ or $SS$ will be open and play a main role in achieving the correct relic density. Then the masses of 
$X_R$, $E_1$, $E_2$ and $N$ are allowed to have sizable deviation from $m_{X_I}$. When $min(m_{Z'},m_S)$ is larger than $m_{X_I}$ and the $X_I X_I \to Z'Z',~SS$ processes
are kinematically forbidden, $min(m_{E_1},~m_{E_2},~m_N,~m_{X_R})$ is required to be closed to $m_{X_I}$ so that 
the correct DM relic density is obtained via their co-annihilation processes.

The $X_I$ has no interactions to the SM quark, and its couplings to the muon lepton and vector-like leptons are constrained by the muon $g-2$ anomaly.
Therefore, the model can easily satisfy the bound from the direct detection of DM. At the LHC, the vector-like leptons are mainly produced via 
electroweak processes,
\bea
&&p~p\to \gamma / Z^* \to E_1\bar{E}_{1,2},~ E_2\bar{E}_{1,2},~N\bar{N},\nonumber\\
&&p~p\to W^* \to E_{1,2} \bar{N},~\bar{E}_{1,2} N,
\eea
then the decay modes include
\beq
E_{1,2} \to \mu X_I, ~~N\to \nu_\mu X_I.
\eeq
If kinematically allowed, the following decay modes will be open,
\beq
E_{1,2}\to \mu X_R,~W N,~~ E_{1,2}\to Z E_{2,1},~~ N\to \nu_\mu X_R.
\eeq

The $2\mu+E_T^{miss}$ event searches at the LHC can impose strong constraints on the vector-like leptons and DM.
The production processes of $2\mu+E_T^{miss}$ in our model are very similar to the electroweak production of charginos and sleptons decaying into final states 
with $2\ell+E_T^{miss}$ analyzed by ATLAS with 139 fb$^{-1}$ integrated luminosity data \cite{1908.08215}.  
Therefore, we will use this analysis to constrain our model, which is implemented in the $\textsf{MadAnalysis5}$ \cite{ma5-1,ma5-2,ma5-3}.
We perform simulations for the samples using \texttt{MG5\_aMC-3.3.2}~\cite{Alwall:2014hca} 
with \texttt{PYTHIA8}~\cite{Torrielli:2010aw} and 
\texttt{Delphes-3.2.0}~\cite{deFavereau:2013fsa}.

If the DM relic density is achieved via the co-annihilation processes of vector-like lepton, the 
mass of vector-like lepton is required to be closed to $m_{X_I}$. As a result, the $\mu$ from the
vector-like lepton decay is too soft to be distinguished at detector, and the scenario can 
easily satisfy the constraints of the direct searches at LHC. Here, we employ the ATLAS analysis of $2\ell+E_T^{miss}$ in Ref. \cite{1908.08215}
to constrain another scenario in which $1<m_{X_R}/m_{X_I}<1.15$ and $m_Z'~(m_S)>m_{X_R}$ is taken, and the co-annihilation processes of $X_R$ can play
a main role in achieving the correct relic density. Thus, the masses of the vector-like leptons
 are allowed to be much larger than $m_{X_I}$.

\begin{figure}
 \epsfig{file=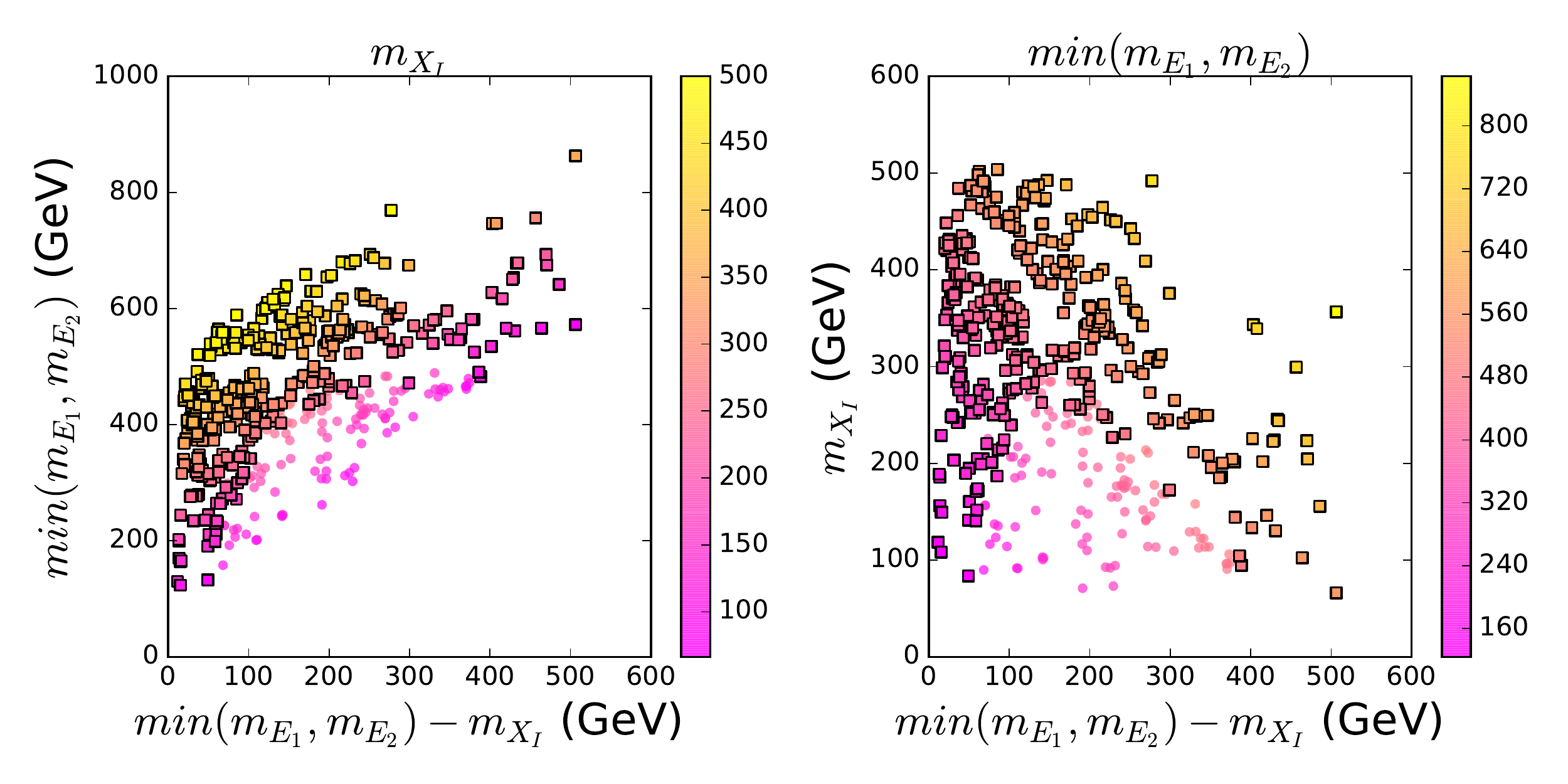,height=8.cm}
\vspace{-0.5cm} \caption{For the scenario of $1<m_{X_R}/m_{X_I}<1.15$ and $m_{Z'}~(m_S)>m_{X_R}$,  all samples satisfy the constraints of "pre-muon $g-2$", 
the diphoton signal data of the 125 GeV Higgs, the muon $g-2$ anomaly, and the DM relic density.
The squares and bullets are allowed and excluded by the direct searches for $2\ell+E^{miss}_T$ at the LHC.}
 \label{figdmxr}
\end{figure}

We impose the constraints of "pre-muon $g-2$", 
the diphoton signal data of the 125 GeV Higgs, the muon $g-2$ anomaly, the DM relic density, and the direct searches for $2\ell+E^{miss}_T$ at the LHC,
and project the surviving samples in Fig. \ref{figdmxr}. From Fig. \ref{figdmxr} we see that the mass of the lightest charged vector-like lepton is allowed 
to be as low as 120 GeV if $min(m_{E_1},~m_{E_2})-m_{X_I}<$ 60 GeV since the muon becomes soft in the region. 
As $min(m_{E_1},~m_{E_2})-m_{X_I}$ increases, the energy of muon becomes large, and the vector-like lepton needs to be large enough
to escape the constraints of direct searches for $2\ell+E^{miss}_T$ at the LHC. For example, $min(m_{E_1},~m_{E_2})$ is favored to
be larger than 500 GeV for $min(m_{E_1},~m_{E_2})-m_{X_I}>$ 300 GeV. The DM mass is allowed to be as low as
100 GeV if $min(m_{E_1},~m_{E_2})-m_{X_I}<$ 60 GeV or $min(m_{E_1},~m_{E_2})-m_{X_I}>$ 400 GeV.

\section{Conclusion}

In this paper we discussed the CDF $W$-mass, the muon $g-2$, and the DM observables in a local $U(1)_{L_\mu-L_\tau}$ model, and obtained the following observations:
(i) The CDF $W$-mass disfavors $m_{E_1}= m_{E_2}={m_N}$ or $s_L=s_R=0$, and favors a large mass splitting 
between $E_1$ and $E_2$ when the differences between $s_L$ and $s_R$ becomes small.
(ii) The muon $g-2$ anomaly can be simultaneously explained for appropriate difference between $s_L$ $(m_{E_1})$ and $s_R$ $(m_{E_2})$, and some regions
are excluded by the diphoton signal data of the 125 GeV Higgs. 
(iii) Combined with the CDF $W$-mass, muon $g-2$ anomaly and other relevant constraints, 
the correct DM relic density is mainly achieved in two different scenarios:
(1) $X_IX_I\to Z'Z',~ SS$ for $m_{Z'}(m_S)<m_{X_I}$ and (2) the co-annihilation processes for $min(m_{E_1},m_{E_2},m_N,m_{X_R})$ closed to $m_{X_I}$.
(iv) The direct searches for $2\ell+E_T^{miss}$ event at the LHC impose strong bounds on the masses of the vector-like leptons and DM as well
as their mass splitting.

\section*{Acknowledgment}
We thank Songtao Liu, Shuyuan Guo, Shiyu Wang, Liangliang Shang, and Yang Zhang for the helpful discussions. This work was supported by the National Natural Science Foundation
of China under grant 11975013.

\begin{appendix}
\section{The $\Pi$ function}
The $\Pi_{XY}(p^2,m_1^2, m_2^2)$ and $\Pi_{XY}(0,m_1^2, m_2^2)$ are given as
\begin{eqnarray}
\Pi_{XY}(p^2,m_1^2, m_2^2)&=&
-{N_c\over 16 \pi^2}\biggl\{
{2\over 3}\biggl(g_{LX}^{f_1f_2}g_{LY}^{f_1f_2}+
g_{RX}^{f_1f_2}g_{RY}^{f_1f_2}\biggr)
\biggl[m_1^2+m_2^2-{p^2\over 3}-
\biggl(A_0(m_1^2)+A_0(m_2^2)\biggr)
\nonumber \\
&&+{m_1^2-m_2^2\over 2 p^2}
\biggl(A_0(m_1^2)-A_0(m_2^2)\biggr)\nonumber\\
&&+{2p^4-p^2(m_1^2+m_2^2)-(m_1^2-m_2^2)^2\over 2 p^2}
B_0(p^2, m_1^2,m_2^2)\biggr]
\nonumber \\
&&
+2m_1m_2\biggl(g_{LX}^{f_1f_2}g_{RY}^{f_1f_2}+
g_{RX}^{f_1f_2}g_{LY}^{f_1f_2}\biggr)B_0(p^2,m_1^2,m_2^2)\biggr\},
\end{eqnarray}
\begin{eqnarray}
\Pi_{XY}(0,m_1^2, m_2^2)&=&
-{N_c\over 16 \pi^2}\biggl\{
{2\over 3}\biggl(g_{LX}^{f_1f_2}g_{LY}^{f_1f_2}+
g_{RX}^{f_1f_2}g_{RY}^{f_1f_2}\biggr)
\biggl[m_1^2+m_2^2 -
\biggl(A_0(m_1^2)+A_0(m_2^2)\biggr)
\nonumber \\
&&-{m_1^2+ m_2^2\over 2 }
B_0(0,m_1^2, m_2^2)
-{(m_1^2-m_2^2)^2\over 2 }
B_0^{'}(0, m_1^2,m_2^2)\biggr]
\nonumber \\
&&
+2m_1m_2\biggl(g_{LX}^{f_1f_2}g_{RY}^{f_1f_2}+
g_{RX}^{f_1f_2}g_{LY}^{f_1f_2}\biggr)B_0(0,m_1^2,m_2^2)\biggr\}.
\end{eqnarray}
Here the coupling constants $g_{LX}^{f_1f_2}$ and $g_{RX}^{f_1f_2}$ are from 
\begin{equation}
{\overline f}_1
\biggl(g_{LX}^{f_1f_2} P_L
+g_{RX}^{f_1f_2} P_R\biggr)\gamma_\mu f_2 X^\mu\,,
\end{equation}
and the $A_0$, $B_0$, and $B_0^{'}$ functions are
\begin{eqnarray}
A_0(m^2)&=&\biggl({4\pi\mu^2\over m^2}\biggr)^\epsilon
\Gamma(1+\epsilon)
\biggl({1\over \epsilon}+1\biggr) m^2 ,
\nonumber \\
B_0(p^2,m_1^2,m_2^2)&=&\biggl({4\pi\mu^2
\over m_2^2}\biggr)^\epsilon\Gamma(1+\epsilon)
\left[{1\over \epsilon} -f_1(p^2,m_1^2,m_2^2)\right] ,\nonumber\\
B_0^{'}(p^2,m_1^2,m_2^2)&=& \frac{\partial}{\partial p^2} B_0(p^2,m_1^2,m_2^2) ,
\end{eqnarray}
and
\begin{equation}
f_1(p^2, m_1^2,m_2^2)=\int_0^1 dx \log\biggl(x+{m_1^2(1-x)-p^2x(1-x)\over m_2^2}
\biggr)\, .
\end{equation}
\end{appendix}

\end{document}